\documentclass[prl, preprint]{revtex4-1} 

\usepackage{amsfonts} 
\usepackage{graphicx}
\usepackage{amsmath, amssymb}
\usepackage{slashed}
\usepackage{braket}
\usepackage{physics}
\usepackage[bottom]{footmisc}
\graphicspath{  {images/} }
\begin{document}


\title{Speeding up the Karatsuba algorithm}

\author{Satish Ramakrishna}
\affiliation{Department of Physics \& Astronomy, \\
Rutgers, the State University of New Jersey, \\
136 Frelinghuysen Road
Piscataway, NJ 08854-8019}
\email{ramakrishna@physics.rutgers.edu} 
\author{Kamesh Aiyer}
\affiliation{Kashi Software Inc, \\
11 Magazine Street,
Cambridge, MA 02139}
\email{kamesh@acm.org} 


\begin{abstract}

The Karatsuba method was the first published attempt to speed up the multiplication of two numbers \cite{Karatsuba}. It is well-known to be of complexity $\sim {\cal O}(n^{\log_2 3})$. There are currently, improved mathematical algorithms, that surpass this algorithm in complexity \cite{Schon, Toom, Cook, Dutt, Furer}.

While it does not seem possible to improve upon the basic mechanism of the Karatsuba technique and we demonstrate why it is the most efficient of its type, it is possible that one can improve its implementation for particular decile ranges of numbers. This article presents an approach to speed up the implementation of the Karatsuba technique, utilizing extra memory to supplement the original method.  While the method is conceptually similar to the ``Method of Four Russians'' technique used to speed up Matrix Multiplications, it applies the concept in a different area.

\end{abstract}

\maketitle 

The Karatsuba algorithm \cite{Karatsuba, Karatsuba2}, an ${\cal O}(n^{\log_2 3})$ technique to multiply two $n$-digit numbers, has been surpassed by newer techniques that are ${\cal O}(n \times \log n \times \log \log n)$ \cite{Schon, Toom, Cook, Dutt, Furer} and ${\cal O}(n \times \log n)$ \cite{Harvey} respectively. However, the simplicity of the algorithm allows improvements that are easily implemented and can be reduced to fewer multiplications, supplemented by look-ups.

\section{The Karatsuba algorithm}

For simplicity, consider multiplying two $n$-digit numbers $x$ and $y$, written as
\begin{eqnarray}
x = x_0 + x_1 10^m \nonumber \\
y = y_0 + y_1 10^m
\end{eqnarray}
where for simplicity,we use $n=2^k$,  $m = \frac{n}{2}$ and work in base-$10$.
The product can be simplified to
\begin{eqnarray}
x . y = x_0 y_0 +   (x_0 y_1+x_1 y_0) \times 10^m + x_1 y_1 10^{2m} \nonumber \\
=x_0 y_0 +  \left( (x_0+x_1)(y_0+y_1) - x_0 y_0 - x_1 y_1 \right) \times 10^m + x_1 y_1 10^{2m} 
\end{eqnarray}
so that the product of the $n$-digit numbers can be reduced to the multiplication of three independent $m$-digit (and occasionally $m+1$-digit) numbers, instead of four $m$-digit numbers. Note that multiplications by $10$ are ignored in the complexity calculations, since they can be reduced to decimal point shifts before additions. Also, as a general principle, the complexity calculation ignores the number of additions, since they are sub-dominant in complexity.
The order of magnitude (``complexity") of the number of separate multiplications to multiply these two numbers can be reduced to the relation
\begin{eqnarray}
{\cal M}(n) = 3 {\cal M}(\frac{n}{2}) + {\cal O} (n) \rightarrow {\cal M}(n) \sim {\cal O}(n^{\log_2 3})
\end{eqnarray}
The Karatsuba technique is sometimes referred to as the ``divide-by-two-and-conquer'' method.

An alternative approach to computing the complexity above is as follows. At every step, the starting number (initially $n=2^s$ digits long) is split into two numbers with half the number of digits. After $s$ steps, it is easy to see that $3^s$ multiplications of single-digit numbers need to be performed. This number, the number of multiplications required, can be written as $3^s=3^{\log_2 n} =n^{\log_2 3}$; hence the above complexity result. This calculation is exact (as explained above) for a number with a power of $2$ as the number of digits.

\section{The improved version}
We generalize the Karatsuba divide-by-two-and-conquer algorithm as follows and write (in base-$B$)
\begin{eqnarray}
x = x_0 + x_1 \times B^m + x_2 \times B^{2m} + ... + x_N \times B^{Nm} \nonumber \\
y = y_0 + y_1 \times B^m + y_2 \times B^{2m} + ... + y_N \times B^{Nm} 
\end{eqnarray}
As can be quickly checked, each of the numbers $x_0,..., x_N, y_0,...,y_N$ are $m$-digits long and $(N+1)m=n$ where $n$ is the total number of digits in $x$ and $y$.

The number of multiplications required to multiply $x$ and $y$, i.e., the order of complexity, is ${\cal M} = (N+1)+{\cal C}(N+1,2) \sim \frac{(N+1)^2}{2} \sim \frac{n^2}{2m^2}$ individual products of $m$-digit (and occasionally $m+1$-digit) numbers. For instance, if $N=1$, as in the usual Karatsuba technique, ${\cal M}=2+1=3$ which is what we use in the order-of-magnitude estimate in Equation (3).

In the Karatsuba technique, the $m (=2)$ digit numbers are further multiplied by the same technique, carrying on recursively till we are reduced to single-digit multiplications. That leads to the recursive complexity calculation noted in Equation (3).

However, note that if we simply pre-computed the individual $m$-digit multiplications and looked up the individual multiplications, we end up with essentially $\sim \frac{n^2}{2m^2}$ lookups rather than actual multiplications. Indeed, lookups take, on average, $1/5$ the time taken for single-digit multiplication (and then we have to multiply by the number of operations ${\cal L} $ required to perform the lookup), hence the  complexity when lookups are added are $\sim \frac{n^2}{10m^2} \times {\cal L} $ in comparison with the Karatsuba method. As we will show below, ${\cal L} \sim 6m$, so that the total complexity of the algorithm is $\sim \frac{n^2}{m} $. If  $m$ were chosen to be a fraction of $n$, i.e., $m = \frac{n}{N+1}$,  the complexity is $\sim (N+1)n$. When compared to the Karatsuba technique, this is much quicker than $n^{1.58}$. This is the main result of this short note.

The lookups of $m$-digit multiplications need to be performed against a table of size $B^m \times B^m$. This lookup, as can be verified by standard binary-search techniques, is (for $B=10$) of complexity $\sim \log_2 (10^{2m}) = 2 m \log_2 10 \sim 6m$. There are some additional additions and subtractions, which add additional (though sub-dominant) complexity $\sim \frac{n}{m}$ as can be easily checked and are detailed in the below example.

Analyzing this further, we could choose to mix and match in two different ways, i.e., 
\begin{enumerate}
\item We could apply $k$ Karatsuba-style divide-by-two-and-conquer steps, then apply the lookup method to look-up $3^k$ pre-calculated products of $\frac{n}{2^k}$ digits or
\item We could use the $(N+1)=2^k$-block technique (break-up into $m$ digit blocks) with  $m=\frac{n}{2^k}$, we'd have to look-up $2^k+{\cal C}(2^k,2)$ products. 
\end{enumerate}
\noindent It is clear that $3^k < 2^k+{\cal C}(2^k,2)$, so the divide-by-two-and-conquer strategy yields fewer lookups - it is the quickest way to speed up the calculation. A graph of the reduced complexity (essentially $3^N (\frac{2}{3})^{N-k}$) achieved this way is plotted in Figure 1 - clearly, cutting the recursion off early is advantageous.

A little reflection will show why divide-and-conquer by $2$ for $k$ times followed by lookup is the most efficient way to carry out the above procedure, in fact it is illustrative for it demonstrates why the Karatsuba technique is the most efficient of the divide-and-conquer techniques. Each time we divide an $n$-digit number into $N+1$ blocks of $m$-digits, we have to (recursively) perform $(N+1)+{\cal C}(N+1,2)$ multiplications. After $k$ such recursions, we are left with $(N+1)^k$ blocks of $\frac{n}{(N+1)^k}$ digits each and have to perform $((N+1)+{\cal C}(N+1,2))^k$ multiplications. At this point, if we look up pre-computed products of numbers of this type, that is a complexity factor of $\sim \frac{n}{(N+1)^k}$. The total number of operations is 
\begin{eqnarray}
\sim ((N+1)+{\cal C}(N+1,2))^k \times \frac{n}{(N+1)^k} = n (1+\frac{N}{2})^k \nonumber
\end{eqnarray}
which is smallest for smallest $N$, i.e., $N=1$. The complexity then matches exactly the complexity of the Karatsuba algorithm.

\begin{figure}[h!]
\caption{Plot of the Efficiency of cutting off Karatsuba early}
\centering
\includegraphics[scale=.52]{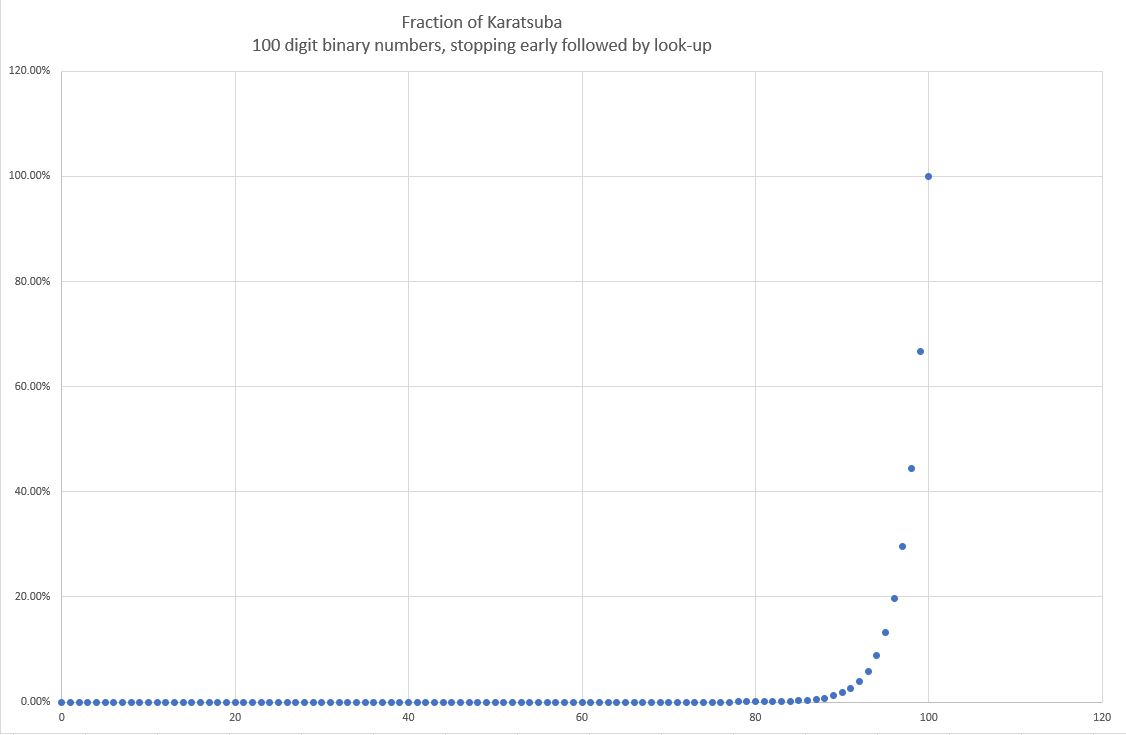}
\end{figure}

\subsection{A numerical example}
The above arithmetic is demonstrated in the case below for $N=4$, where we have used $n=5m$,
\begin{eqnarray}
x = x_0+x_1 B^m+x_2 B^{2m}+x_3 B^{3m} + x_4 B^{4m} \nonumber \\
y =y_0+x_1 B^m+y_2 B^{2m}+y_3 B^{3m} + y_4 B^{4m}
\end{eqnarray}
which leads to the product
\begin{eqnarray}
x.y = B^0 x_0 y_0 + B^m \left( (x_0+x_1)(y_0+y_1) - x_0 y_0 -x_1 y_1 \right) \nonumber \\
+ B^{2m} \left( (x_0+x_2)(y_0+y_2)-x_0 y_0 -x_2 y_2+x_1 y_1\right) \nonumber \\
+ B^{3m} \left( (x_0+x_3)(y_0+y_3) - x_0 y_0 - x_3 y_3 +(x_1+x_2)(y_1+y_2) - x_1 y_1 - x_2 y_2 \right) \nonumber \\
+ B^{4m} \left( (x_0+x_4)(y_0+y_4) - x_0 y_0 - x_4 y_4 +(x_1+x_3)(y_1+y_3) - x_1 y_1 - x_3 y_3 + x_2 y_2 \right) \nonumber \\
+ B^{5m} \left( (x_1+x_4)(y_1+y_4) - x_1 y_1 - x_4 y_4 +(x_2+x_3)(y_2+y_3) - x_2 y_2 - x_3 y_3 \right) \nonumber \\
+ B^{6m} \left( (x_2+x_4)(y_2+y_4)-x_2 y_2 -x_4 y_4+x_3 y_3 \right) \nonumber \\
+ B^{7m} \left( (x_3+x_4)(y_3+y_4) - x_3 y_3 -x_4 y_4  \right)+ B^{8m} \left( x_4 y_4 \right)
\end{eqnarray}
As can be observed, this expression has $5+{\cal C}(5,2) =15$ independent products that can be pre-computed, i.e.., the $5$ simple products $x_0 y_0, x_1 y_1, x_2 y_2, x_3 y_3, x_4 y_4$ and the $10$ combination products $(x_0+x_1)(y_0+y_1), (x_0+x_2)(y_0+y_2), (x_0+x_3)(y_0+y_3), (x_0+x_4)(y_0+y_4) , (x_1+x_2)(y_1+y_2), (x_1+x_3)(y_1+y_3),  (x_1+x_4)(y_1+y_4), (x_2+x_3)(y_2+y_3), (x_2+x_4)(y_2+y_4), (x_3+x_4)(y_3+y_4)$. If these products are found in a pre-computed table of $m$ and $(m+1)$-digit numbers, we would not need any multiplications at all, just 15 lookups, for any $m$, with $n=5m$. 

We would need to perform additions and subtractions, of course and there are 34 of them in the above example - this number of elementary operations depends, however, only upon $\frac{n}{m}$.

\subsection{Memory Requirements}

Typical RSA encryption algorithms use $\sim 1000$-digit base-$10$ composite numbers that are the product of five-hundred-digit primes. If one were to attack the problem by pre-computing keys, i.e., pre-multiplying pairs of five-hundred-digit primes $(n=500 \sim 2^9)$ and storing the results of multiplying all possible $6$-digit numbers $(m=6 \sim 2^3)$, one has a complexity $\sim \frac{n^2}{m} = 85 n \sim 42,500$, which is worse than the new $\sim n \log_2 n \sim 4500$ complexity  \cite{Harvey}, albeit the fact that the newer approach also has multipliers, which we have not accounted for. If we use the hybrid method (Karatsuba followed by look-up of 6-digit products),  the complexity is $\sim 3^6 \times 6 \sim 4200$, which is arguably much better (no pre-factors missing) than even the $n \log_2 n$ algorithms. We would need to store $\sim 10^{12}$ twelve-digit numbers, roughly 20 TB of memory, which is a reasonable size.

\section{Conclusion}

This paper presents a rapid pre-computed approach to speeding up multiplications. Though one needs to pre-compute and store all possible $m$-digit multiplications, one can compute the products of two integers with number of digits equal to any integer times $m$ in time proportional to the number of digits (times the afore-mentioned integer). Memory is cheaper than CPU-speed, so this is a method that can be exploited in other (for instance signal-processing) situations to speed up intensive calculations too.

Useful conversations are acknowledged with Dr. B. Kumar. As this paper was being prepared, an article about using pre-stored calculations was released, where the Eratosthenes sieve was sped up in calculation complexity \cite{Harald}.

\end{document}